\documentstyle[12pt,titlepage,cites]{article}
 \newcommand{\beq}{\begin{equation}}
 \newcommand{\eeq}{\end{equation}}
 \newcommand{\beqa}{\begin{eqnarray*}}
 \newcommand{\eeqa}{\end{eqnarray*}}

 \def\pabl#1#2{\frac{\partial {#1}}{\partial { #2}}}

\begin{document}
\begin{center}
{\Large Entangled Polymer Rings in 2D and Confinement}
\vspace{5cm}

M.Otto and T.A.Vilgis\\[12pt]
Max-Planck-Institut f\"ur Polymerforschung\\ 
Postfach 3148, D-55021 Mainz, F.R.G.\\
\end{center}
\vspace{10ex}
\noindent
\vspace{2ex}
36.20 Macromolecules and polymeric molecules
  
%
\newpage
\setcounter{equation}{0}
\section{Introduction}
The statistical mechanics of entangled polymers, i.e., 
polymer chains under topological constraints is a generally unsolved 
problem. The main difficulty is to specify 
distinct topological states of the polymer chain. Closed polymers, or
polymer loops appear to be a much simpler system as they are either 
linked 
(with themselves or with one another) or unlinked. Linear
chains, however, can always be disentangled. On a shorter time scale 
than the disentanglement time though,
it seems justified to define topological states "on the average", using the 
same formalism like for polymer loops.

Mathematically, the problem of specifying topological states of polymer loops 
is equivalent to the classification problem for knots and links 
\cite{kauffman:93}. 
Since the mid-eighties considerable progress has been made 
following Jones \cite{jones:85} as various new knot polynomials have 
been discovered. (For a review on knots see 
\cite{kauffman:93}). 

For an analytical theory of the polymer entanglement problem, 
the algebraic form of these invariants is not 
suitable (see section V for new perspectives). They are generally expressed in one, 
two or three
variables which appear in the defining relations (known as skein relations). 
There is no immediate relation of these variables to the polymer 
conformation, and consequently there is no reasonable
way how to couple the knot invariant to a statistical weight for a given
polymer conformation. 
Algebraic knot theory seems to be applicable for 
the theory of entangled directed polymers
\cite{nechaev:95}, and certainly does
so in computer simulations (e.g. 
\cite{volod:74}\cite{volod:75}\cite{koniaris:91}\cite{tsurusaki:95}).

The degrees of freedom appearing in the statistical weight of given 
polymer conformation
are usually 
expressed in terms of segment positions
${\bf r}(s)$ which is a mapping $[0, N]\rightarrow R^d$ from the contour 
variable
$s$ to $d$-dimensional space \cite{doi:86}. 
For $d=3$, one invariant showing an explicit dependance on these variables is the 
the so called Gauss invariant 
for two given closed loops $C_\alpha$ and $C_\beta$
parametrized by ${\bf r}_\alpha(s)$, ${\bf r}_\beta(s)$
\begin{equation}
\label{1.1}
\Phi(C_\alpha, C_\beta)=\frac{1}{4\pi}\oint_{C_\alpha}ds\oint_{C_\beta}ds'
\dot{{\bf r}}_\alpha (s)\wedge\dot{{\bf r}}_\beta (s')\cdot
\frac{{\bf r}_\alpha (s)-{\bf r}_\beta (s')}
{|{\bf r}_\alpha (s)-{\bf r}_\beta (s')|^3}
\enspace,
\end{equation}
which is invariant with respect to continuous deformations of the loops,
was first used by Edwards to discuss entangled polymer loops 
\cite{edwards:67}\cite{edwards:68}. It is also called the Gaussian 
linking number (sometimes also the winding number).
As Edwards already noted, in $d=3$ the Gauss invariant 
 does not uniquely specify a given link.
There is, in fact, an infinite series of higher order link invariants that appear quite naturally 
in a perturbative expansion of a field-theoretical representation of 
the generalized Jones polynomial \cite{witten:89}\cite{guada:93}.
For special cases such as a random walk winding around an obstacle of 
infinite length, rigorous results can be obtained using the Gauss invariant,
as it has been discussed in detail by Wiegel \cite{wiegel:86}.

As even the Gauss invariant is difficult to handle mathematically for 
a rigorous treatment of the three-dimensional entanglement problem,
many
mean field type arguments have been used \cite{doi:86} for
a rough characterization of the topological states. Tube models
for entangled polymer melts are the most prominent approaches,
which seem to lead to contradictory conclusions for open linear
polymer chains and closed ring polymers. Therefore more detailed
knowledge about the topological states of polymer systems
is needed. In general, this seems to be a difficult task,
and to overcome mathematical and conceptual difficulties simple
and model type situations must be studied, to learn about more
complex ones. 
Only recently more detailed situations are under consideration by using 
the path integral approach and 
the Gaussian linking number constraint \cite{brereton:95}.

To give an example that shows the complexity of the structure
of the theory we mention the following development:
The "easiest" topological arrangement of closed polymer rings is
the non-concatenated melt of rings, since all winding numbers
between different rings are zero.
Scaling arguments for the typical size of a ring, $R \propto N^{\nu}$, 
have  been put forward \cite{cates:86}, giving an estimate of 
$\nu = 2/5$. 
This result seems to be in rough agreement with
computer simulation \cite{vilgisweyersberg:94} 
(see also \cite{geyler:88}\cite{wittmer:95}). 
The more detailled analytical many chain theory in \cite{brereton:95}
supported the scaling result, although the
theory is more involved than simple scaling.

An important step forward 
to formulate the problem by field theoretic methods 
was made by Brereton and Shah \cite{brereton:80}. A test loop 
in the melt was considered that is entangled with many other chains.
The resulting theory can be mapped to Euclidean electrodynamics in 
$3=2+1$ dimensions coupled to a $O(n)$ $\phi^4$ theory 
for the conformation 
of a self-avoiding walk (in the limit $n\rightarrow 0$ \cite{degennes:72}
\cite{descloizeaux:90}). 
It is the basis for work by 
Nechaev and Rostiashvili \cite{nechaev:93} in two dimensions.

As a matter of fact, for $d=2$
the Gauss invariant becomes rigorous for "simple", 
i.e. non-self-intersecting loops. It reads
\begin{equation}
\label{1.2}
G_i (C)=\frac{1}{2\pi}\oint_{C}ds\dot{{\bf r}}(s)\cdot \nabla
\left(\ln|{\bf r}(s)-{\bf r}_i|\right)\wedge
{\mbox{\boldmath$\eta$ }}
\end{equation}
 Here the vector ${\bf r}(s)$ represents the segment positions 
of the polymer loop
in the plane. 
In Eq.(\ref{1.1}),
the role of the polymers entangled with the loop
is taken formally by obstacles at the positions ${\bf r}_i$.  ${\mbox{\boldmath$\eta$ }}$ is a unit vector perpendicular to plane \cite{nechaev:93}. 
$G_i (C)$ is also called the winding number of the loop.
Eq.(\ref{1.2}) may also be expressed in terms of a Cauchy integral in the 
complex plane and is related to the oriented area of the loop in the plane 
\cite{brereton:87}. For so-called "complex" loops, i.e. loops with points of
self-intersection, special care is needed as 
the invariant Eq.(\ref{1.2}) might give zero although the
loops are entangled (see
Rostiashvili, Nechaev, and Vilgis \cite{rost:93}).

The theoretical 
basis of this paper is the field theory by Brereton et al.
\cite{brereton:80}.
It was studied by Nechaev and Rostiashvili \cite{nechaev:93}\cite{rost:93}
in order to discuss the behavior of a 
polymer loop in an array
of randomly distributed parallel 
line obstacles \cite{nechaev:93}\cite{rost:93} whose spatial distribution 
is quenched.
 A first order transition for critical length $N_c$ for the 
polymer loop was found when the quenched 
average over the 
winding number distribution was taken. It has been interpreted as a collapse
transition (for $N>N_c$) to an octopus conformation resembling a randomly branched polymer 
with no self-interaction
 with an end-to-end vector scaling of the size $R \propto
N^{\frac{1}{4}}$, where $N$ is the total length of the ring \cite{zimm:49}
\cite{lubensky:78}\cite{lubensky:79}.

In fact, in the work below it will be shown explicitly 
(within the approximation made in
\cite{nechaev:93}) that 
the conjectured scaling behavior is valid even for annealed 
disorder with respect to the spatial distribution of
obstacles. This assumption is reasonable if the obstacles are supposed 
to represent other polymers entangled with a given loop, and was 
also made in the original theory by Brereton et al. \cite{brereton:80}.
We will obtain the same free energy (in the mean-field approximation) as
Nechaev et al.\cite{nechaev:93} - yet in a physically more 
transparent form - and consequently, the same transition behavior.

We will exploit the fact that the interaction due to the topological constraint
is a so-called "area law", i.e. it is proportional to the area enclosed by the 
loop.
 This is an exact result known in quantum field 
theory in the context of Wilson 
loops and the confinement problem for abelian gauge fields 
(see e.g. \cite{zinn-justin:93}).
Therefore, the area of the loop 
is equal to first order 
to the effective potential found in \cite{nechaev:93} and is
dependant on the conformational fields and topological quantities. 

The paper is organized
as follows: In section II we clarify that the 
effective interaction is proportional to the area being the order 
parameter of the problem. 
In section III, we show on the level of polymer field theory 
how the area appears in the effective interaction and how it 
depends on the conformational fields (in the mean field approximation).
The critical 
behavior of the order 
parameter as a function of topological parameters is discussed. 
A new upper bound for the range of stability for 
the mean field solution (in the replica-symmetric case) 
is found.
In section IV, the conjectured scaling like 
for randomly branched polymers is 
shown,  and in section V a brief outlook on the complete, i.e. $d=3$ entanglement problem is 
given.

\section{The area as the order parameter}

In the work of Nechaev and Rostiashvili \cite{nechaev:93} it is not evident 
at first sight that the first order phase transition caused by 
topological disorder corresponds to a classical
collapse transition, similarly
to the case of a polymer immersed
in bad solvent \cite{descloizeaux:90}. We expect indeed 
structural differences between a collapsed chain in bad solvent and a 
"collapsed ring" under topological constraints.
 
In order to understand better the nature of the phase transition, 
it is desirable to introduce 
an order parameter that fits to the problem, which is in our case the area of
the 2$d$ projection of a {\em simple} loop. Following Cardy \cite{cardy:94} we use the 
covariant expression
\begin{equation}
\label{2.1}
A=\int d^2{\bf x}\int d^2{\bf x'}\langle A_\mu ({\bf x})A_\nu ({\bf x'})
\rangle J_\mu ({\bf x})J_\nu ({\bf x'}).
\end{equation}
where ${\bf x}=(x,y)$ is a vector in 2-dimensional Euclidean space.
Variables $J_\mu$ are the polymer current densities or 
tangent vector densities $J_\mu ({\bf x})=\int_0^N ds\dot{r}_\mu(s)
\delta({\bf x}-{\bf r}(s))$ where $N$ is the chemical length of the chain.
The gauge fields $A_\mu$ are of the U(1) type and their correlator is gauge
dependant. Using the gauge $A_1=0$ it reads as 
$\langle A_\mu ({\bf x})A_\nu ({\bf x'})
\rangle=-\frac{1}{2}\delta_{\mu 0}\delta_{\nu 0} |y-y'|\delta(x-x')$, and
it becomes obvious that the r.h.s of Eq.(\ref{2.1}) is indeed an area.

When using expression Eq.(\ref{2.1}) for the definition of
the area some care is needed.  The above equation is valid 
only for non-self-intersecting loops which will be 
the scope of the present treatment. In the case of self-intersections,
negative area contributions may cancel positive ones giving a total zero 
area (see above the discussion following Eq.(\ref{1.2})).

We next recall the incorporation of constraint Eq.(\ref{1.2}) 
in the partition sum for the loop \cite{nechaev:93}. For a fixed number 
of obstacles $c$ enclosed by the loop the partition function is given by:
\begin{eqnarray}
\label{2.2}
Z(c)&=&\int{\cal D}{\bf r}(s)\delta({\bf r}(N)-{\bf r}(0))
\delta\left(c-\oint_{C}ds\dot{{\bf r}}(s)\cdot {\bf A}\right)\\
&&\exp\left(
-\frac{1}{l^2}\oint ds\dot{{\bf r}}^2(s)-\frac{a^2}{2}\oint ds\oint ds'
\delta({\bf r}(s)-{\bf r}(s'))
\right)\nonumber
\end{eqnarray}
$l$ is the Kuhn segment length, and $a^2$ is the 2$d$ excluded volume.
From Eq.(\ref{1.2}) we see that the gauge field ${\bf A}$ 
is given by $\sum_{i=1}^N\nabla
\left(\ln|{\bf r}(s)-{\bf r}_i|\right)\wedge{\mbox{\boldmath$\eta$ }}$ so that 
the delta function fixes the winding number in terms of 
the 2D Gauss invariant. 
It satisfies 
$\nabla\cdot {\bf A}=0$ and $\nabla\wedge{\bf A}={\mbox{\boldmath$\eta$}}\left(
\varphi({\bf r})-\varphi_0\right)$ where 
$\varphi({\bf r})=\sum_i\delta({\bf r}-{\bf r}_i)$ and $\varphi_0$ is the mean 
density of obstacles in the $xy$-plane.

In contrast to \cite{nechaev:93} 
we suppose that first, 
the spatial distribution of obstacles is annealed whereas second,
the distribution of winding numbers $c$, i.e. obstacles enclosed by the loop
is quenched. The first assumption is reasonable
 if the obstacles are to represent
other polymers (of infinite length) entangled with the loop, which is an important feature 
of a model
that projects the original three-dimensional entanglement problem to two 
dimensions.

The second assumption is clear from the fact 
that once a given winding number $c$ is 
fixed for the loop, it should remain fixed in the process of averaging 
over both the conformations of the loop and the positions of obstacles. 
As a consequence, 
we take the {\em annealed} average 
over the spatial distribution of obstacles to be the gaussian
\begin{equation}
\label{2.3}
P\left(\varphi({\bf r})\right)
\sim
\exp\left(-\frac{1}{2\varphi_0}\int d^2{\bf x}
(\varphi({\bf r})-\varphi_0)^2\right) 
\sim \exp\left(-\frac{1}{2\varphi_0}\int d^2{\bf x}
(\nabla\wedge{\bf A})^2\right).
\end{equation}
The {\em quenched} distribution 
of the number of obstacles $c$  is assumed to be
a gaussian with mean $c_0$ and dispersion $\Delta_c$,
\begin{equation}
\label{wind}
P(c)\sim \exp\left(-\frac{(c-c_0)^2}{2\Delta_c}\right)
\end{equation}
and is used to average the free energy.

 The winding number 
constraint in the partition sum is expressed by a Fourier transform introducing 
the variable $g$, a chemical potential conjugate to $c$:
\begin{equation}
\label{ntog}
\delta\left(c-\oint_{C}ds\dot{{\bf r}}(s)\cdot {\bf A}\right)=
\int \frac{dg}{2\pi}e^{igc-ig\oint_{C}ds\dot{{\bf r}}(s)\cdot {\bf A}}
\end{equation}
 The distribution of 
winding numbers $P(c)$ may then be transformed into a distribution $P(g)$ 
for the chemical potential \cite{nechaev:93}. 
For later purposes it is crucial to note that the $g^2$ averaged over 
$P(g)$ is 
\beq
\label{gaverage}
[ g^2]_g=
\frac{1}{\Delta_c}\left(1-\frac{c^2_0}{\Delta_c}\right).
\eeq
The partition function is now expressed in terms of $g$. After averaging 
over the the spatial distribution of obstacles, it reads:
\begin{eqnarray}
\label{2.4}
\langle Z(g)\rangle_{\bf A}&=&{\cal N}\int{\cal D}{\bf A}\delta(\nabla\cdot{\bf A})
\int{\cal D}{\bf r}(s)\delta({\bf r}(N)-{\bf r}(0))\nonumber\\
&&\exp\left(
-\frac{1}{2\varphi_0}\int d^2{\bf x}
(\nabla\wedge{\bf A})^2-
ig\oint_{C}ds\dot{{\bf r}}(s)\cdot {\bf A}\right.\nonumber\\
&&\left.-\frac{1}{l^2}\oint ds\dot{{\bf r}}^2(s)-\frac{a^2}{2}\oint ds\oint ds'
\delta({\bf r}(s)-{\bf r}(s'))
\right)
\end{eqnarray}
${\cal N}$ is a normalization factor for the average over the gauge fields .

Carrying out the integral over the gauge fields ${\bf A}$, one obtains:
\begin{eqnarray}
\label{2.5}
\langle Z(g)\rangle_{\bf A}&=&\int{\cal D}{\bf r}(s)\delta({\bf r}(N)-{\bf r}(0))\nonumber\\
&&\exp\left(
-
\frac{\varphi_0 g^2}{2}
\int d^2{\bf x}\int d^2{\bf x'}\langle A_\mu ({\bf x})A_\nu ({\bf x'})
\rangle J_\mu ({\bf x})J_\nu ({\bf x'})
\right.\nonumber\\
&&\left.-\frac{1}{l^2}\oint ds\dot{{\bf r}}^2(s)-\frac{a^2}{2}\oint ds\oint ds'
\delta({\bf r}(s)-{\bf r}(s'))
\right)
\end{eqnarray}
The resulting term in the exponential is proportional to the area of the 
loop.
In fact, the interaction reads as
\begin{equation}
\label{2.6}
\beta H_{\rm int}=\frac{\varphi_0 g^2}{2}A
\end{equation}
If we replace $g^2$ by its mean value Eq.(\ref{gaverage}) we obtain:
\begin{equation}
\label{2.7}
\beta H_{\rm int}=\frac{\varphi_0}{2\Delta_c}\left(1-\frac{c_0^2}{\Delta_c}
\right)A.
\end{equation}
The approximation considered in \cite{nechaev:93} is limited to range of values 
for $\varphi_0$ and $\Delta_c$ which require the factor in front of $A$ to be
always positive (as to further restrictions on the set of values for these
parameters
where a mean field solution is valid, see the end of section III). Consequently, in order to minimize its energy the loop tends
to collapse, decreasing its area. 

\section{The area as a collective variable and polymer field theory}

We now introduce the area explicitly as collective variable in the 
partition sum using
$1=\int dA \delta(A-\hat{A})$ where $\hat{A}$ is given by Eq.(\ref{2.1}).
After transforming the delta function and some standard 
manipulations we then have:
\begin{eqnarray}
\label{3.1}
\langle Z(g)\rangle_{\bf A}&=&{\cal N}\int dA\int d\alpha
\int{\cal D}{\bf r}(s)\delta({\bf r}(N)-{\bf r}(0))\nonumber\\
&&\exp\left(
i\alpha A
-(i\alpha+
\frac{\varphi_0 g^2}{2})
\int d^2{\bf x}\int d^2{\bf x'}\langle A_\mu ({\bf x})A_\nu ({\bf x'})
\rangle J_\mu ({\bf x})J_\nu ({\bf x'})
\right.\nonumber\\
&&\left.-\frac{1}{l^2}\oint ds\dot{{\bf r}}^2(s)-\frac{a^2}{2}\oint ds\oint ds'
\delta({\bf r}(s)-{\bf r}(s'))
\right)
\end{eqnarray}
Let us define $i\tilde{\alpha}=i\alpha+
\frac{\varphi_0 g^2}{2}$ for a moment, and
express $\alpha$ in terms of $\tilde{\alpha}$,
 and make the change $\alpha\rightarrow\tilde{\alpha}$ in the
integration variable. This procedure looks strange at first glance,
because the new integration variable $\tilde{\alpha}$ becomes now
complex. Below we show that this is not a serious problem
for the purpose in this paper. After these manipulations we obtain:
\begin{eqnarray}
\label{3.2}
\langle Z(g)\rangle_{\bf A}&=&{\cal N}\int dA\int d\tilde{\alpha}
\int{\cal D}{\bf r}(s)\delta({\bf r}(N)-{\bf r}(0))\nonumber\\
&&\exp\left(-\frac{\varphi_0 g^2}{2}A+
i\tilde{\alpha}A
-i\tilde{\alpha}
\int d^2{\bf x}\int d^2{\bf x'}\langle A_\mu ({\bf x})A_\nu ({\bf x'})
\rangle J_\mu ({\bf x})J_\nu ({\bf x'})
\right.\nonumber\\
&&\left.-\frac{1}{l^2}\oint ds\dot{{\bf r}}^2(s)-\frac{a^2}{2}\oint ds\oint ds'
\delta({\bf r}(s)-{\bf r}(s'))
\right)
\end{eqnarray}
In the next step, the term depending on the gauge field correlator is expressed
 in terms of a gaussian integration:
\begin{eqnarray}
\label{3.3}
\langle Z(g)\rangle_{\bf A}&=&{\cal N}
\int{\cal D}{\bf A}\delta(\nabla\cdot{\bf A})
\int dA\int d\tilde{\alpha}
\int{\cal D}{\bf r}(s)\delta({\bf r}(N)-{\bf r}(0))\nonumber\\
&&\exp\left(
-\frac{1}{2}\int d^2{\bf x}
(\nabla\wedge{\bf A})^2
-ie
\int d^2{\bf x}A_\mu ({\bf x})
 J_\mu ({\bf x})
-\frac{\varphi_0 g^2}{2}A+
i\tilde{\alpha}A
\right.\nonumber\\
&&\left.-\frac{1}{l^2}\oint ds\dot{{\bf r}}^2(s)-\frac{a^2}{2}\oint ds\oint ds'
\delta({\bf r}(s)-{\bf r}(s'))
\right)
\end{eqnarray}
$e$ is a shorthand notation for $\sqrt{2i\tilde{\alpha}}$ 
and is the "coupling constant" of the analoguous Wilson loop problem, well 
known in quantum field theory \cite{zinn-justin:93}. We note again that
$e$ is complex, but it will turn out below that this is not a problem.
The partition sum has now a structure similar to the original formulation 
considered in \cite{nechaev:93} and is suitable for a 
field theoretic treatment. 
Note that the interaction $\frac{\varphi_0 g^2}{2}A$ has been 
completely separated
from the conformational average and the average over the 
distribution of obstacles. The area is related to the conformation 
directly
only via 
the coupling constant $e$ or $\tilde{\alpha}$ respectively.
In fact, the problem will be first examined 
for a given realization of these variables. 
In the last step, the parameter $\tilde{\alpha}$ will be 
eliminated to give back the dependance of the area on the conformation 
of the loop.

To proceed further, we consider only consider 
the functional intregation over the positions ${\bf r}(s)$, i.e. the 
partition function
\beqa
\label{Aeg}
Z(e;[{\bf A}])&=&{\cal N}
\int{\cal D}{\bf r}(s)\delta({\bf r}(N)-{\bf r}(0))\exp\left(
-\frac{1}{l^2}\oint ds\dot{{\bf r}}^2(s)
\right.\nonumber\\
&&\left.-\frac{a^2}{2}\oint ds\oint ds'
\delta({\bf r}(s)-{\bf r}(s'))
-ie
\int d^2{\bf x}A_\mu ({\bf x})
 J_\mu ({\bf x})
\right)
\eeqa
The partition function 
Eq.(\ref{Aeg}) describes the statistics of the loop for given external 
"magnetic" field ${\bf A}$. 
It is 
formally the same partition function as $Z(g)$ in Eq.(\ref{2.4}), and 
can be 
treated in terms of
the $n$ vector
$\phi^4$ theory in the limit $n\rightarrow 0$ \cite{degennes:72}
according to \cite{brereton:80}\cite{nechaev:93}. 
To obtain the field theory, the following standard steps have to be 
carried out. First,
the two-dimensional excluded volume interaction is expressed in terms of
a gaussian average over a
pseudopotential \cite{doi:86}.
Then, one has to consider the Green's function of the loop for a given 
realization of the gauge field ${\bf A}$ and the pseudopotential for 
the excluded volume. The Green's function is expressed in terms of a Gaussian 
field theory. The averages over the pseudopotential and the gauge field 
lead to consider an n-fold
replicated field theory
 (see \cite{nechaev:93} for technical details). When the average over 
the pseudopotential is carried out, one finally obtains
\begin{eqnarray}
\label{3.4}
Z_n(e;[{\bf A}])
&=&
\prod_{i=1}^n\left(
\int {\cal D}{\bf \phi}_i\int {\cal D}{\bf \phi}^\ast_i\right)\exp\left(
-\int d^3 {\bf R} {\cal H}[{\bf A}, \phi_i, \phi_i^*]
\right),
\end{eqnarray}
with
\begin{eqnarray}
\label{3.5}
{\cal H}[{\bf A}, \phi_i, \phi_i^*]&=&\sum_{i=1}^n\phi_i\left(
m^2-\frac{l^2}{4}(\nabla_\perp-ie{\bf A})^2-\frac{l^2}{2}
\nabla_\parallel^2
\right)\phi_i^\ast
+\frac{La^2}{4}\sum_{i,j=1}^n\phi_i\phi_i^\ast\phi_j\phi_j^\ast\nonumber\\
\end{eqnarray}
The fields $\phi_i$ 
are replica fields for polymer loops. The model
has been embedded into 3-dimensional space, so the excluded volume term is 
the embedded 2-dimensional one with $L$ being the mean size of the polymer
in the $z$ direction.

Now, the average over the obstacle distribution is taken as
follows:
\beq
\langle Z_n (e;[{\bf A}])\rangle_{\bf A} =
{\cal N}\int {\cal D}{\bf A}\delta(\nabla\cdot{\bf A})Z_n  (e;[{\bf A}])
\exp\left(
-\frac{1}{2}\int d^2{\bf x}(\nabla\wedge {\bf A})^2
\right)
\eeq
The gauge fields are integrated out following \cite{nechaev:93}.
To simplify the algebra, the Landau gauge is used.
One obtains an effective action with the one-loop correction being given 
by:
\begin{equation}
\label{3.7}
{\cal H}_{1-{\rm loop}}=\int \frac{d^2{\bf k}}{(2\pi)^2}\log\left(
k^2+Q\sum_i\phi_i\phi_i^\ast
\right)
\end{equation}
with $Q=\frac{l^2}{2}e^2=i\tilde{\alpha}l^2$. As $Q$ is complex, 
a complex logarithm in Eq.(\ref{3.7}) has to be considered. It is
easily shown that by  restricting 
the analysis to one Riemannian sheet, the integral can be evaluated 
straight-forwardly under the assumption
that the $\phi_i$ are constant in space. This assumption is consistent
with the result that the topologically restricted chain forms a dense
object with very small density fluctuations. This corresponds indeed
to the assumption of $\phi_i \approx$ constant. 

In \cite{nechaev:93} the replica-symmetric case is studied because only
in this case the effective potential approximation can be used.
(For the details of solving integral Eq.(\ref{3.7}) and the renormalization
procedure we refer the reader to \cite{nechaev:93}).

In this case, we now approximate 
$\sum_i\phi_i\phi_i^*=n\phi\phi^*$.
After renormalization according to \cite{nechaev:93}
one obtains:
\begin{equation}
\label{3.8}
\langle Z_n (e=e(\tilde{\alpha});[{\bf A}])\rangle_{\bf A}
=
\exp\left(
-\int d^3 {\bf R} {\cal L}_{eff}
\right)
\end{equation}
with an effective Lagrangian
\begin{eqnarray}
\label{3.9}
{\cal L}_{eff}&=&in\tilde{\alpha}\left(
-\frac{l^2}{4\pi}|\phi|^2\ln(\frac{|\phi|^2}{M^2})+\frac{l^2}{2\pi}|\phi|^2
\right)
+n(m^2-La^2M^2)|\phi|^2+n\frac{La^2}{4}|\phi|^4\nonumber\\
\end{eqnarray}
$M$ is an arbitrary subtraction point appearing due to the renormalization 
procedure.  
Because the fields $\phi$ are now constant in space, the integration in 
Eq.(\ref{3.9}) gives simply a constant volume factor $V$.
From Eq.(\ref{3.8}) one obtains
the contribution to the free energy as a function of 
$\tilde{\alpha}$ which is conjugate to the area. 
Devided by the system volume, it is given using the standard 
formula:
\begin{eqnarray}
\label{replica}
f(\tilde{\alpha})=\frac{{\cal F}(\tilde{\alpha})}{V}
&=&\frac{1}{V}\pabl{}{n}\langle 
Z_n (e=e(\tilde{\alpha});[{\bf A}])\rangle_{\bf A}
|_{n=0}\nonumber\\
&=&i\tilde{\alpha}\left(
-\frac{l^2}{4\pi}|\phi|^2\ln(\frac{|\phi|^2}{M^2})+\frac{l^2}{2\pi}|\phi|^2
\right)
+(m^2-La^2M^2)|\phi|^2+\frac{La^2}{4}|\phi|^4\nonumber\\
\end{eqnarray}
The next step is to transform back from $\tilde{\alpha}$
to the area $A$. This is done by a Legendre transfrom (or by a Fourier 
transform of the partition function).
Finally one has to add the area term $\frac{\varphi_0 g^2}{2}A$ 
to the free energy 
which yields the partition function averaged over the distribution of 
obstacles:
\beq
\label{3.10}
\langle Z(g)\rangle_{\bf A}=\int dA\delta\left(
A+\frac{l^2}{4\pi}V|\phi|^2\ln(\frac{|\phi|^2}{M^2})
-\frac{l^2}{2\pi}V|\phi|^2\right)
e^{- Vf(A,g)} 
\eeq
with the free energy density
\beq
\label{3.11}
f(A,g)=\frac{\varphi_0 g^2}{2V}A
+(m^2-La^2M^2)|\phi|^2+\frac{La^2}{4}|\phi|^4.
\eeq
The set of Eq.s (\ref{3.10}) and (\ref{3.11}) is the fundamental result
of this paper.
The free energy density $f(A,g)$ is indeed
the area law plus the renormalized action for the
self-avoiding walk loop. 
Integrating over the area gives back the result of
\cite{nechaev:93} for the free energy density in terms of conformational fields 
and $g$ only.
Averaging the free energy density 
in Eq.(\ref{3.11}) over the distribution of winding numbers 
using Eq.(\ref{gaverage}) one finally obtains
\beq
\label{3.11a}
f(A)=[f(A,g)]_g=\frac{\varphi_0}{2V\Delta_c}\left(1-\frac{c^2_0}{\Delta_c}\right)A
+(m^2-La^2M^2)|\phi|^2+\frac{La^2}{4}|\phi|^4
\eeq
As a consequence of Eq.(\ref{3.11a}), the essential result
we have obtained here is the dependance of 
the area on the fields $\phi$ in the mean field approximation expressed
in the delta function of Eq.(\ref{3.10}). 
\begin{equation}
\label{3.12}
A=-\frac{l^2}{4\pi}V|\phi|^2\ln(\frac{|\phi|^2}{M^2})+\frac{l^2}{2\pi}V|\phi|^2
\end{equation}
Introducing the segment density $\rho=|\phi|^2$ and choosing $M^2=L^{-3}$ 
we then obtain:
\begin{equation}
\label{3.13}
A=V\frac{l^2}{2\pi}\left(\rho-\frac{1}{2}\rho\ln(\rho L^3)\right).
\end{equation}
Let us now investigate the behavior of the area when the collapse transition
occurs. It has been studied in detail in \cite{nechaev:93}, so we just 
mention the results. In fact, the collapse transition
takes place
at the critical length $N_c$ which is given by (see 
\cite{nechaev:93}, but 
with the factor $\frac{1}{32\pi}$ replaced by $\frac{1}{8\pi}$):
\begin{equation}
\label{3.14}
\frac{1}{N_c}=\frac{l^2}{8\pi}\frac{\varphi_0}{\Delta_c}\left(
1-\frac{c_0^2}{\Delta_c}\right)\ln\left[
\left(
\frac{L}{a}
\right)^2
\frac{l^2}{4\pi}\frac{\varphi_0}{\Delta_c}\left(
1-\frac{c_0^2}{\Delta_c}\right)
\right]
\end{equation}
As $1/N_c$ is always positive, the condition
\begin{equation}
\label{3.14b}
\left(
\frac{L}{a}
\right)^2
\frac{l^2}{4\pi}\frac{\varphi_0}{\Delta_c}\left(
1-\frac{c_0^2}{\Delta_c}\right)
>1
\end{equation}
follows for the stability of the mean field solution.
At $N=N_c$ the segment density is:
\begin{equation}
\label{3.15}
\rho_c=\frac{1}{La^2}\frac{l^2}{4\pi}\frac{\varphi_0}{\Delta_c}\left(
1-\frac{c_0^2}{\Delta_c}\right).
\end{equation}
Thus the critical area is:
\begin{equation}
\label{3.16}
A_c=\frac{V}{2La^2}\left(\frac{l^2}{2\pi}\right)^2
\frac{\varphi_0}{\Delta_c}\left(
1-\frac{c_0^2}{\Delta_c}\right)
\left(1-\frac{1}{2}\ln
\left[
\left(
\frac{L}{a}
\right)^2
\frac{l^2}{4\pi}\frac{\varphi_0}{\Delta_c}\left(
1-\frac{c_0^2}{\Delta_c}\right)
\right]
\right).
\end{equation}
$A_c$ essentially depends on the topological parameters $c_0$, the mean 
winding number, $\Delta_c$, the dispersion of the winding number distribution, 
and $\varphi_0$ the mean density of obstacles in the plane. As $A_c$ should
remain non-negative we obtain
a new upper bound in addition to the inequality 
Eq.(\ref{3.14b}):
\begin{equation}
\label{3.17}
\mbox{e}^2\geq\left(
\frac{L}{a}
\right)^2
\frac{l^2}{4\pi}\frac{\varphi_0}{\Delta_c}\left(
1-\frac{c_0^2}{\Delta_c}\right)
>1
\end{equation}
where e$=2.714...$ is Euler's constant (e is not to be confounded with
the coupling constant $e$ defined earlier). Eq.(\ref{3.17})
indicates that at a certain value of the mean density of obstacles 
$\varphi_0$ the mean field solution will break down. However, 
the specific value
obtained 
for the upper bound is a result of the mean field expression for the 
critical area $A_c$, Eq.(\ref{3.16}), and should not be taken as 
a quantitative, but a qualitative result. Nechaev et al. have stressed that the mean field 
approximation is valid in the vicinity of the boundary curve 
 defined by the lower bound inequality Eq.(\ref{3.14b})
\cite{nechaev:93} (see FIG.(1)). The present approach using the area of the loop as the 
order parameter and giving the new upper bound gives strong support of 
this result that the mean field solution is restricted to a small neigborhood 
above the boundary curve.

\section{Final result and discussion}
It has been argued in \cite{nechaev:93} that the collapsed phase can be 
identified with a randomly branched polymer. 
Here we give explicit support of this idea.
Consider Eq.(\ref{3.13}) and substitute $\rho=\frac{N}{V}$ in the mean field
approximation where $N\geq N_c$, i.e. above the critical length. We
obtain:
\begin{equation}
\label{4.1}
A=\frac{l^2}{2\pi}N\left(1-\frac{1}{2}\ln(N)+...\right).
\end{equation}
Taking the terms in parentheses as the first powers of an exponential, one 
finds:
\begin{equation}
\label{4.2}
A=\frac{l^2}{2\pi}N(N^{-\frac{1}{2}})=\frac{l^2}{2\pi}N^\frac{1}{2}.
\end{equation}
Exploiting the result of Cardy \cite{cardy:94} for SAWs in $d=2$ 
that $\langle A\rangle 
\sim \langle R^2\rangle$, one obtains for $R=(\langle R^2\rangle)^\frac{1}{2}$:
\begin{equation}
\label{4.3}
R\sim lN^\frac{1}{4}
\end{equation}
This is the scaling behavior for randomly branched polymers without excluded
volume interaction. It is therefore very likely that the collapsed phase of 
the loop corresponds to a randomly branched polymer. 

Note that the 
result of Eq.(\ref{4.2}) corresponds to the free part of the 
free energy ${\cal F}$ in Eq.(\ref{3.11}). i.e. without excluded volume. 
This is valid as a first approximation because the area $A$ is not directly 
coupled to the density $\rho=|\phi|^2$ at the level of the free energy, and 
the excluded volume is not renormalized by the topological interactions in 
the  
mean field approximation. 
Moreover, a Landau expansion in terms of segment density $\rho$ and tangent vector
density variables $j_\mu$ indicates
 that the $\rho$ and $j_\mu$ decouple 
at first order because $k_\mu j_\mu ({\bf k})=0$,
 while interactions occur only at higher order. \cite{brereton:95}

As a consequence, the two-dimensional excluded volume must 
still be taken into account. That yields the well known
$d=2$ branched polymer scaling \cite{lubensky:78,lubensky:79} for the area or 
the mean sqare end-to-end vector respectively, i.e. $R^2\sim lN^\frac{5}{4}$.


\section{Outlook to the three-dimensional problem}
Finally, let us point out some
possible future perspectives for the $d=3$ entanglement problem 
by coming to the mathematical difficulty of 
finding a correct knot invariant. While the known knot invariants in their 
algebraic form seem to be only of limited use in the polymer context, the work of Witten 
\cite{witten:89}
showing an equivalence of the Jones polynomial (actually a more general one) 
and (in general non-abelian) 
Chern-Simons field theory has brought the knot problem closer to physics 
again. Witten showed that the expectation value of Wilson lines averaged over
a Chern-Simons action functional integral 
gives a knot invariant for framed links. This invariant 
can also be reproduced perturbatively 
giving the sqare of the Gauss invariant as
a first approximation and higher order knot invariants \cite{cotta:90}\cite{guada:90}
\cite{guada:93}. 

In addition it has been shown recently that the invariants appearing in 
Chern-Simons perturbation theory are intimately related to 
so-called Vassiliev invariants (see e.g. \cite{alvarez:95} and the 
literature quoted therein).

These results
may open new ways of solving the polymer entanglement problem by a 
"topological perturbation theory". It might give a range of validity for
using the Gauss invariant for ensemble of random walk chains or rings.

\newpage
FIG. 1: The phase diagram of the collapse transition. The shaded area 
bounded by the solid curve corresponds to the 
collapsed state of the loop as 
obtained in [22]. The new upper bound in Eq.(32) gives rise to 
the new boundary curve (dashed line).
\end{document}